\documentstyle[11pt,preprint,aps]{revtex}
\begin{document}
\draft

\title{Hermiticity and the Cohomology Condition in Topological
Yang-Mills Theory\thanks{This work is supported in part by funds
provided by the U. S. Department of Energy (D.O.E.) under contract
\#DE--AC02--76ER03069.}}
\author{Jean-Guy Demers\footnote{jgdemers@mitlns.mit.edu}}
\address{Center for Theoretical Physics, \\ Laboratory for Nuclear Science \\
and Department of Physics\\
Massachusetts Institute of Technology, Cambridge, MA 02139}
\maketitle

\begin{abstract}
The symmetries of the topological Yang-Mills theory are studied in the
Hamiltonian formalism and the generators of the twisted N=2 superPoincar\'e
algebra are explicitly constructed.  Noting that the twisted Lorentz
generators do not generate the Lorentz symmetry of the theory, we relate
the two by extracting from the latter the twisted version of the
internal SU(2) generator.
The hermiticity properties of the various
generators are also considered throughout,
and the boost generators are found to be
non-hermitian.  We then recover the
BRST cohomology condition on physical states from representation theory
arguments.
\end{abstract}

\vspace{\fill}
\noindent CTP\#2233
\hfill September 1993 \break
\vspace{-24pt}
\newpage
\section{INTRODUCTION}
\label{sec:1}

In recent years, much attention has been devoted to the study of topological
field theories \cite{Birming}.  Because these theories have no
local dynamics, their correlation functions depend only on the global features
of the target space. An important example is given by
the topological Yang-Mills (TYM) theory, which
 was used
to obtain the Donaldson invariants
for smooth 4-manifolds\cite{Witten}.  Shortly after TYM was introduced,
it was shown \cite{Brooks,Singer} that
it can also be obtained by BRST gauge fixing the
topological symmetry ($\delta A_\alpha^a = \theta_\alpha^a$, with
$\theta_\alpha^a$ arbitrary) of either zero or the topological action
$S=\int \, d^4 x \
F^{\mu\nu} \, \tilde{F}_{\mu\nu}$. Under an appropriate choice of
gauge parameters, the resulting action is identical to the one
introduced in \cite{Witten} and is given by:
\begin{eqnarray}
{S} &=& \int_M \: Tr \biggl\{  {1\over 4} F_{\alpha \beta}
F^{\alpha \beta}-
{1\over 2} D_\alpha \phi D^\alpha \lambda - i \eta D_\alpha \psi ^\alpha
+ i D_\alpha \psi_\beta \chi^{\alpha \beta} \nonumber\\
& &\mbox{}\hskip .7in
- {i \over 8} \phi \left[
\chi_{\alpha \beta},\chi^{\alpha \beta} \right] -{i \over 2} \lambda
\left[ \psi_{\alpha},\psi^{\alpha} \right]-{i \over 2} \phi
\left[ \eta,\eta \right]-{1\over 8} \left( \left[ \phi,\lambda \right]
\right)^2 \biggl\}\nonumber\\
&\equiv & \int_M\: {\cal L}.
\label{RLa}
\end{eqnarray}
Here, all fields are Lie algebra valued and transform according to the adjoint
representation of the gauge group, which is taken to be compact and
semi-simple.  The covariant derivative is
$D_\alpha= \nabla_\alpha+[A_\alpha,\  ]$, where $\nabla_\alpha$
is the covariant derivative with respect to the diffeomorphisms on the
curved manifold $M$ of metric $g_{\mu\nu}$.
The gauge field $A_\alpha$ and the scalars $\lambda$ and $\phi$
are bosonic  while $\eta$,
$\psi_\alpha$ and   $\chi_{\alpha\beta}$ are all
anticommuting and respectively scalar, vector and self-dual tensor fields
$\left(\chi_{\alpha\beta}=
{1\over 2}\varepsilon_{\alpha\beta}{}^{\mu\nu}\chi_{\mu\nu}\right)$.
Note that in this version (\ref{RLa}) still possesses the usual
(non-topological) Yang-Mills symmetry.

When (\ref{RLa}) was
introduced, its intimate relation with $N=2$ super Yang-Mills (SYM) was
already noticed \cite{Witten}. In fact, formal representations of the
Donaldson polynomials have also been obtained in SYM\cite{BK}.
In Euclidean space-time, the latter theory enjoys the
Lorentz symmetry $SO(4)$ (isomorphic to $SU_L(2) \bigotimes
SU_R(2)$) as well as the internal global $SU_I(2)$ symmetry.
If one ``twists'' this
symmetry by replacing $SU_L(2)$ by the diagonal sum of $SU_L(2)$ and
$SU_I(2)$, $SU_{L^{\prime}} (2)$,
the rotation group then becomes $SU_{L^{\prime}} (2)
\bigotimes SU_R (2)$ and the resulting theory is just (\ref{RLa}).
Through this procedure the original supersymmetry generators are also
transformed and the Lorentz scalar supercharge thus obtained is identified as
the BRST charge. The twist procedure has also been used to obtain
extended $(N=2)$ TYM theories\cite{Y}. Furthermore, TYM has also been obtained
via
the use of Killing spinors in $N=2$ conformal supergravity\cite{KR}; in this
case, a ``local'' version of the twisting procedure is implemented by embedding
the $SU(2)$ connection in the Lorentz spin connection.

In this paper, we will detail the twisting of the $N=2$ supersymmetry
(Section~\ref{sec:2}) and explicitly construct the various generators while
studying their hermiticity properties (Section~\ref{sec:3}).
We will argue that after
twisting, the internal symmetry generators are transformed into a
useful and hitherto unappreciated symmetry of (\ref{RLa}).  It will also
be shown that the boost generators are not hermitian.  This will be
used in Section~\ref{sec:4} to discuss the following issue.  Despite their
connection through twisting, TYM and SYM theories differ in that the
former does not support any local excitations. When TYM is considered through
the BRST
construction, it is found that the only states in the cohomology
of the BRST charge are
those with vanishing energy \cite{Witten}.
Among other things, this absence of local excitations complicates any attempt
to use
topological field theories in a description of quantum gravity, a
possibility suggested by
the natural general covariance of
these theories.
It is hence usually thought that one must first
establish a mechanism to break the topological symmetry.
As a consequence, we find it compelling to study more closely
the relation between SYM and TYM.
Within the context of twisting and without
appealing to the BRST derivation of (\ref{RLa}),
we will propose an explanation, based on representation theory arguments,
of why TYM is
indeed free of local excitations. As we are only interested in the
details of the
canonical quantization of the theory, such as its hermiticity properties and
spectrum, we will work on flat manifolds.
Our concluding remarks are contained in Section~\ref{sec:5}.

\section{TWISTED N=2 SUPERSYMMETRY ALGEBRA}
\label{sec:2}

Our starting point is the $N=2$ superPoincar\'e algebra
(without central charge) \cite{West,Rocek}:
\begin{mathletters}
\begin{eqnarray}
[P_\alpha, P_\beta] &=& 0 ,
\hspace{0.5in}
[ P_\mu, J_{\alpha\beta} ] = i \, g_{\mu [\alpha} \ P_{\beta]} , \\
\relax [J_{\alpha\beta}, J_{\mu\nu}] &=&  i g_{\alpha[\mu}\ J_{\nu]
\beta}
+ i g_{\beta[\mu}\ J_{\alpha|\nu]}  ,\\
\relax [P_{\mu}, Q_{Ai}]  &=&  0 = [P_\mu, \bar{Q}_{\dot{A} j}] ,\\
\relax [Q_{Ai}, J_{\alpha\beta}] &=& (\sigma_{\alpha\beta})_A {}^{B} Q_{Bi} ,
\hspace{0.5in}
[\bar{Q}^{\dot{A}} {}_{j}, J_{\alpha\beta}]
= (\bar{\sigma}_{\alpha\beta})^{\dot{A}} {}_{\dot{B}}
\bar{Q}^{\dot{B}} {}_{j} ,\\
\relax \{ Q_{Ai}, \bar{Q} {}_{\dot{B}} {}^{j} \} &=& 2 \delta {}_{i}
{}^{j} P_\alpha (\sigma^{\alpha})_{A \dot{B}} ,\\
\relax \left\{ Q_{Ai}, Q_{Bj} \right\} &=& 0 = \{ \bar{Q}_{\dot{A}} {}^{i},
\bar{Q}_{\dot{B}} {}^{j} \}  ,\\
\relax [T^i{}_j, J_{\alpha\beta}]  &=&  0 = [T^i{}_j,P_\alpha] ,\\
\relax [T^i{}_j,T^k{}_l]&=&\frac{1}{2} \left(\delta^k_j\
T^i{}_l-\delta^i{}_l \ T^k{}_j\right) ,\\
\relax [T^i{}_j,Q_{Ak}]&=&-\frac{1}{2} \left(\delta^i_k\
Q_{Aj}-\frac{1}{2}\delta^i_j\
Q_{Ak} \right) ,\\
\relax  [T^i{}_j,\bar{Q}_{\dot{A} k}]&=&-\frac{1}{2} \left(\delta^i_k\ \bar{Q}_
{\dot{A}j}-\frac{1}{2}\delta^i_j\ \bar{Q}_{\dot{A}k} \right) .
\end{eqnarray}
\label{kkk}
\end{mathletters}
Our convention closely follows Ref.~\cite{Sohnius}.
Bracketed indices are to be antisymmetrized, ignoring the ones just
preceeding a
vertical bar (thus $g_{\beta[\mu} J_{\alpha|\nu]}\equiv
g_{\beta\mu}J_{\alpha\nu}-
g_{\beta\nu}J_{\alpha\mu}$).
Greek letters denote
Lorentz indices, with $P_\alpha$ and $J_{\alpha\beta}$ standing for
translation and Lorentz generators respectively.  Capital latin
letters are two-spinor indices with undotted ones referring to $SU_R(2)$ and
dotted ones to $SU_L(2)$.  Raising and lowering these indices is done with the
help of the antisymmetric matrices $ \varepsilon_{AB},\varepsilon^{AB},
\varepsilon_{\dot{A} \dot{B}}$ and $\varepsilon^{\dot{A} \dot{B}}$.
They are given by: $ \varepsilon_{12}=-\varepsilon^{12}=
\varepsilon_{\dot{1} \dot{2}} =-\varepsilon^{\dot{1} \dot{2}}=-1$ and act on
Weyl spinors as:\ $ \bar{\psi}_{\dot{A}}=\varepsilon_{\dot{A}\dot{B}}
\bar{\psi}^{\dot{B}} , \chi^{A}=\varepsilon^{AB} \chi_{B}$.
The internal indices are $i,j...$ (in
subsequent sections, these symbols will be used as spatial components of
Lorentz indices); $T^i{}_j$ is traceless and generate $SU_I(2)$.
The metric $g_{\alpha\beta}$ is euclidean ($=\delta_{\alpha\beta}$)
whereas $\sigma_\alpha = (-i, \sigma_{j}),
\bar{\sigma}_{\alpha} = (i, \sigma_{j})$ where $\sigma_i$ are the usual
Pauli matrices. Similarly to the Minkowski case, we define
 $\sigma_{\alpha\beta}=\frac{i}{4}(\sigma_\alpha
\bar{\sigma}_\beta-\sigma_\beta \bar{\sigma}_\alpha)$ and $
\bar{\sigma}_{\alpha\beta}=\frac{i}{4}(\bar{\sigma}_\alpha
\sigma_\beta-\bar{\sigma}_\beta \sigma_\alpha)$.

We now perform the twisting of this algebra. Replacing $SU_L(2)$ by the
diagonal
sum of $SU_L(2)$ and $SU_I(2)$ translates into the identification of the
internal indices with left handed Weyl spinor indices, leading to:
\begin{mathletters}
\label{twist}
\begin{eqnarray}
Q_{Ai} &\rightarrow& Q_{A\dot{C}}
= a_1 (\sigma^\alpha)_{A\dot{C}} \bar{Q}_\alpha , \\
\bar{Q}^{\dot{B}} {}_{j} &\rightarrow& \bar{Q}^{\dot{B}} {}_{\dot{D}}
= a_2 \ \delta^{\dot{B}}_{\,\dot{D}} \ Q_{+}
\ a_3 (\bar{\sigma}^{\mu\nu})^{\dot{B}} {}_{\dot{D}} \ S_{\mu\nu} ,\\
T^i{}_j &\rightarrow& T^{\dot{A}}{}_{\dot{B}}=a_4\
(\bar{\sigma}^{\mu\nu})^{\dot{A}}{}_{\dot{B}}\ R_{\mu\nu} ,
\end{eqnarray}
\end{mathletters}
where on the RHS, the twisted quantities are expressed in terms of
their Lorentz components: $\bar{Q}_\alpha$ is
a vector, $Q$ a scalar, whereas $S_{\mu\nu}$ and $R_{\mu\nu}$ are self-dual
tensors; $a_i$'s are
arbitrary constants. At this stage, these constants could be absorbed
in the definition of the
generators, but they will be useful in the next section, as we will use
already known expressions for $Q$ and $\bar{Q}_\alpha$.
$Q,\bar{Q}_\alpha$ and $S_{\mu\nu}$ are Grassman odd, whereas $R_{\mu\nu}$
is Grassman even. Note that vectorial Grassman charges are also known
to exist in
non-critical string theory\cite{B}.
The relations (\ref{twist}) can be
inverted:
\begin{mathletters}
\begin{eqnarray}
\bar{Q}_\alpha &=&
\frac{1}{2a_1} (\sigma_\alpha)^{\dot{C}A} Q_{A\dot{C}} , \\
Q &=& \frac{1}{2a_2} \ \delta^{\dot{D}}_{\,\dot{B}} \
\bar{Q}^{\dot{B}} {}_{\dot{D}} , \\
S_{\mu\nu} &=& \frac{1}{2a_3}
(\bar{\sigma}_{\mu\nu})^{\dot{B}} {}_{\dot{D}}
\ \bar{Q}^{\dot{D}} {}_{\dot{B}} ,\\
R_{\mu\nu} &=& \frac{1}{2a_4}
(\bar{\sigma}_{\mu\nu})^{\dot{A}} {}_{\dot{B}}
\ T^{\dot{B}} {}_{\dot{A}} ,
\end{eqnarray}
\end{mathletters}
where we have made use of the identity:
\begin{equation}
\hbox{Tr~}
\left( \bar{\sigma}_{\alpha\beta} \ \bar{\sigma}_{\mu\nu} \right)
= \frac{1}{2} \delta_{\alpha[\mu}\ \delta_{\nu]\beta}
+ \frac{1}{2} \epsilon_{\alpha\beta\mu\nu}.
\end{equation}
Under the twisting (\ref{twist}), the superPoincar\'e algebra (\ref{kkk})
is transformed
into:
\begin{mathletters}
\label{algebra}
\begin{eqnarray}
\relax [P_\alpha, P_\beta] &=& 0 , \hspace{1in} [P_\mu, J_{\alpha\beta}]
= i \delta_{\mu[\alpha}\ P_{\beta]} \label{alga} , \\
\relax [J_{\alpha\beta}, J_{\mu\nu}] &=& i \delta_{\alpha[\mu}\ J_{\nu]
\beta}
+ i \delta_{\beta[\mu}\ J_{|\alpha\nu]} \label{algb} , \\
\relax [P_\beta, Q] &=& [P_\beta, \bar{Q}_\alpha] = [P_\beta,
S_{\mu\nu}]
= 0 \label{algc} , \\
\relax [Q, J_{\alpha\beta}] &=& \frac{a_3}{a_2} S_{\alpha\beta}
\label{algd} , \\
\relax [\bar{Q}_\mu, J_{\alpha\beta}] &=& \frac{i}{2}
\left( \delta_{\mu[\alpha}\ S_{\beta]\nu} - \delta_{\nu[\alpha}\ S_{\beta]\mu}
\right) \label{alge} , \\
\relax [S_{\mu\nu}, J_{\alpha\beta}] &=&
\frac{Q}{4} \frac{a_2}{a_3} \left( \delta_{\alpha[\mu} \ \delta_{\nu] \beta }
+ \epsilon_{\alpha\beta\mu\nu} \right)
+ \frac{i}{2} \left( \delta_{\mu [ \alpha} \ S_{\beta ] \nu}
- \delta_{\nu [ \alpha } \ S_{\beta]\mu} \right) \label{algf} , \\
\relax \{ Q, \bar{Q}_\alpha \} &=& -\frac{1}{a_1 a_2} P_{\alpha}
\label{algg} , \\
\relax \{ \bar{Q}_\alpha, S_{\mu\nu} \} &=& -\frac{i}{2 a_1 a_3}
\left( \delta_{\alpha[\mu}\ P_{\nu]}+\epsilon_{\alpha\mu\nu\beta} P^\beta
\right) \label{algh} , \\
\relax \{ \bar{Q}_\alpha, \bar{Q}_\beta\} &=& \{ Q, Q \}
= \{ S_{\mu\nu}, S_{\alpha\beta} \} = 0
\label{algi} ,\\
\relax [Q, R_{\mu\nu}] &=& \frac{a_3}{4a_2a_4} S_{\mu\nu}
\label{algj}  ,\\
\relax [R_{\mu\nu}, S_{\alpha\beta}] &=&
-\frac{a_2}{16a_3a_4} Q \left( \delta_{\alpha[\mu} \ \delta_{\nu] \beta }
+ \epsilon_{\alpha\beta\mu\nu} \right)
- \frac{i}{8a_4} \left( \delta_{\mu [ \alpha} \ S_{\beta ] \nu}
- \delta_{\nu [ \alpha } \ S_{\beta]\mu} \right) \label{algk} , \\
\relax [\bar{Q}_\alpha, R_{\mu\nu}] &=&-\frac{i}{8a_4}
\left( \delta_{\alpha[\mu}\ \bar{Q}_{\nu]} +\varepsilon_{\mu\nu\alpha\beta}
\ \bar{Q}^\beta \right) \label{algl} , \\
\relax [R_{\mu\nu}, R_{\alpha\beta}] &=&
- \frac{i}{4a_4} \left( \delta_{\mu [ \alpha} \ R_{\beta ] \nu}
- \delta_{\nu [ \alpha } \ R_{\beta]\mu} \right) \label{algm} , \\
\relax [R_{\mu\nu},J_{\alpha\beta}]&=&0=[R_{\mu\nu},P_\beta] .\label{algn}
\end{eqnarray}
\end{mathletters}
The existence of (\ref{algebra}) was conjectured in \cite{BK,Ogievetsky}.
The Poincar\'e sector of the algebra Eqs.~(\ref{alga} --- \ref{algb})
is of course left unchanged by the twisting. This would suggest that for the
twisted theory, $J_{\alpha\beta}$ also generate Lorentz rotations. However,
a look at (\ref{algebra}) reveals that the fermionic charges, as well as
$R_{\mu\nu}$, do not transform in the expected way
(e.g. $Q$ does not transform as a scalar). In the following section,
we will examine how the algebra (\ref{algebra}) is realized in TYM, and
identify the correct Lorentz generators.

\section{THE ALGEBRA REALIZED}
\label{sec:3}

In order to study the twisted $N=2$ superPoincar\'e symmetries of (\ref{RLa}),
we make use of
Noether's theorem in its Lagrangian form. Under a symmetry
transformation, the variation of the Lagrangian density is
a total derivative $\delta {\cal L}= \partial_\mu \Lambda^\mu$ and
using the equations of motion, the current
$J^\mu= \sum_{fields \Phi} \delta{\Phi}{{\partial\cal{L}}\over
{\partial(\partial_\mu\Phi)}}-\Lambda^\mu$
is conserved \cite{Jackiw}.
The simplest of the symmetries is the invariance under
translation,
for which
$\delta {\cal L}= {a^\mu}{\partial_\mu} \cal L$
with $a^\mu$ a constant infinitesimal parameter.
The corresponding form of the energy-momentum tensor
is given by:
\begin{eqnarray}
\theta^{\alpha \gamma}&=& F^\gamma{}_\mu F^{\alpha \mu}-
{1\over 2} D^\alpha \phi \:D^\gamma \lambda-{1 \over 2} D^\alpha
\lambda \:D^\gamma \phi+i D^\gamma \psi^\alpha \ \eta+iD^\gamma
\psi_\mu\
\chi^{\alpha \mu}
\nonumber\\
&& \hskip .1in +D_\mu A^\gamma\
F^{\alpha \mu}+ A^\gamma J^\alpha-g^{\alpha \gamma}{\cal L}
\label{EMT}
\end{eqnarray}
with
\begin{equation}
J_\alpha={1\over 2}\left[ \phi, D_\alpha \lambda \right]+
{1\over 2}\left[ \lambda, D_\alpha \phi \right]-
{i}\left[ \psi_\alpha, \eta \right]-
{i}\left[ \psi^\mu, \chi_{\alpha\mu} \right],
\label{jalpha}
\end{equation}
and where ${\cal L}$ is the Lagrangian density given in (\ref{RLa}).
The conservation of this tensor,
$\partial _ \alpha \theta ^{\alpha \gamma}=0$,
gives rise to the energy and momentum generators:
\begin{eqnarray}
P_0&= & \int d^3 x\: \biggl\{ {1\over 2} \left( F_{0i} F_{0i}-
\tilde{F}_{0i}\tilde{F}_{0i}\right) - {1\over 2}D_0\phi\: D_0\lambda
+{1\over 2}D_i\phi\: D_i\lambda- i \varepsilon_{ijk} D_j \psi_k\: \chi_i
\nonumber\\
&& \hskip .6in  -i \psi_0 D_i\chi_i+i \eta D_i\psi_i
+{i\over 2}\phi \left[ \chi_i, \chi_i \right]
+{i\over 2}\lambda \left[ \psi_i, \psi_i \right] \nonumber\\
&& \hskip .6in  +{i\over 2}\lambda \left[ \psi_0, \psi_0 \right]
+{i\over 2}\phi \left[ \eta, \eta \right]
+{1\over 8}\left( \left[\phi,\lambda\right]\right)^2
-A_0G \biggl\} ,
\label{P0}
\end{eqnarray}
\begin{equation}
P_i=\int d^3x\: \left\{ F_{0k} F_{ik} - {1\over 2}D_0\lambda\: D_i\phi
-{1\over 2}D_0\phi\: D_i\lambda + iD_i\psi_j\: \chi_j +i D_i \eta\: \psi_0 -
A_i\: G \right\} ,
\label{Pi}
\end{equation}
where $\chi_i\equiv\chi_{0i}$.
Here and in the rest, integrations are over ``spatial'' coordinates, with
traces understood. We also ignore ordering ambiguities.
In Eq. (\ref{P0}), $A_0$  should be viewed as the Lagrange
multiplier
which imposes the generalized Gauss law constraint $G\equiv D_iF_{0i}-J_0
\approx 0$.
The Hamiltonian (\ref{P0}) can equally be obtained by Legendre
transforming (\ref{RLa}).
In order to compute the algebra of these charges, we first identify the
various momenta of (\ref{RLa}), and impose on them the appropriate equal-time
canonical commutators:
\begin{eqnarray}
P_{\psi_i^a}&=&i\chi_i^a, \hskip 1.99in
\{\chi_i^a(x),\psi_j^b(y)\}=-\delta^{ab}
\delta_{ij}\delta(x-y), \nonumber\\
P_{\psi_0^a}&=&i\eta^a , \hskip 2.01in \{\psi_0^a(x),\eta^b(y)\}=-
\delta^{ab}\delta(x-y), \nonumber\\
P_{A_i^a}&=&F^a_{0i}, \hskip 2.0in [A_i^a(x),F^b_{0j}(y)]=i
\delta^{ab}\delta_{ij}\delta(x-y), \nonumber\\
P_{\phi^a}&=&-{1\over 2} (D_0\lambda)^a, \hskip 1in
[\phi^a(x),-{1\over 2}(D_0\lambda)^b(y)]
=i\delta^{ab}\delta(x-y), \nonumber\\
P_{\lambda^a}&=&-{1\over 2} (D_0\phi)^a, \hskip 1in
[\lambda^a(x),-{1\over 2}(D_0\phi)^b(y)]=i\delta^{ab}\delta(x-y),
\label{CaS}
\end{eqnarray}
where x and y denote here space coordinates. Making use of these commutators,
$P_0$ and $P_i$ are found to correctly translate the
fields, and when commuted among themselves yield:
\begin{equation}
[P_i,P_j]=0, \hspace{1in}[P_0,P_j]=i \int d^3x \:\partial_j A_0(x) G(x).
\label{PiPi}
\end{equation}
As with various forthcoming commutators,
we find that because of
the remnant  Yang-Mills symmetry in the action (\ref{RLa}),
the algebra (\ref{algebra}) is only realized on physical states,
annihilated by the constraint $G$.

We now wish to study the hermiticity properties of our generators.
We take for adjoint assignments:
\begin{eqnarray}
 A_\alpha^\dagger &=& A_\alpha,  \nonumber\\
 \psi_i^\dagger &=& -\chi_i, \nonumber\\
\psi_0 ^\dagger &=& \eta , \nonumber \\
\phi^\dagger&=&\lambda.
\label{adj}
\end{eqnarray}
Despite its non-covariance, this choice is natural for various reasons.
In order for the field theory to be
well defined, $P_0$ should be hermitian and it is under (\ref{adj}).
Moreover, as is shown below,
this choice also leads  to a semi-positive definite spectrum for $P_0$,
in analogy with SYM theory. Also, $P_i$ and the Lagrangian (\ref{RLa}) are
equally hermitian with
this prescription. Note that because of the peculiarity of the self-duality
operation in euclidean metric\cite{Nash} $(\chi_{\alpha\beta}=
{1 \over 2\lambda}
\varepsilon_{\alpha\beta\mu\nu} \chi^{\mu\nu}$ with $\lambda =1,i$
for euclidean and Minkowskian metrics respectively), we require $\varepsilon_
{\alpha\beta\mu\nu}$ to change sign when taking the adjoint. Given that the
presence of $\varepsilon_{\alpha\beta\mu\nu}$ in the various generators
has its origin in the self-dualtiy of $\chi_{\alpha\beta}$, this prescription
in effect reproduce the study of hermiticity in Minkowskian metric.
An alternative road would be to study the Lagrangian (\ref{RLa}) in Minkowski
spacetime. The symmetry generators would then, up to signs, be the same as
the ones presented
here for euclidean metric. The algebra of the generators in that case would be
a Wick rotated version of (\ref{algebra}), obtained by the change:
$\delta_{\mu\nu} \rightarrow \eta_{\mu\nu} (\equiv diag(-1,1,1,1)),\
\sigma^\alpha \rightarrow
(1,\sigma_i),\  {\bar \sigma}^\alpha \rightarrow (-1,\sigma_i),\
\varepsilon_{\alpha\beta\mu\nu} \rightarrow {1 \over i} \varepsilon_
{\alpha\beta\mu\nu}$.

Under Lorentz transformations, the variation of the fields is
$\delta\Phi=\omega^{\mu\nu}x_\mu\partial_\nu\Phi$
where $\omega^{\mu\nu}$ is an infinitesimal antisymmetric parameter.
The corresponding currents:
\begin{eqnarray}
M^{\alpha\beta\gamma}&= & \biggl[x^\beta \biggl\{
-{1\over2}D^\alpha\lambda\:D^\gamma\phi
-{1\over2}D^\alpha\phi\:D^\gamma\lambda
+F^\gamma{}_\mu F^{\alpha\mu}
+iD^\gamma\eta\:\psi^\alpha
+iD^\gamma\psi_\mu\:\chi^{\alpha\mu}\nonumber\\
&&
+{A^\gamma\over 2}[\lambda, {D^\alpha\phi}]
+{A^\gamma\over 2}[\phi, {D^\alpha\lambda}]
-iA^\gamma [\eta, \psi^\alpha]
-iA^\gamma [\psi_\mu, \chi^{\alpha\mu}]
-g^{\alpha\gamma}\cal{L}\biggl\}\nonumber\\
&&
+D_\mu ( x^\beta A^\gamma)\: F^{\alpha\mu}+i\psi^\gamma\chi^{\alpha\beta}
\biggl]-\biggl[ \beta\leftrightarrow\gamma\biggl],
\end{eqnarray}
are conserved $(\partial_\alpha M^{\alpha\beta\gamma}=0)$ and lead to
the constants of motion associated with boosts and rotations:
\begin{equation}
M_{0i}=x_0P_i-\int d^3x \:x_i {\cal P}_0+i\int d^3x\: \psi_i \eta,
\label{M0i}
\end{equation}
\begin{equation}
M_{kj}=\int d^3x\:\left\{\left(x_k {{\cal P}_j}-i \psi_k \chi_j\right) -\left(k
\leftrightarrow j \right)\right\},
\label{Mkj}
\end{equation}
where ${\cal P}_0$ is the energy density, as integrated in Eq.(\ref{P0})
and similarly for ${\cal P}_j$ from Eq.(\ref{Pi}).
One can readily check, using (\ref{adj}),  that $M_{kj}$ is hermitian
but that $M_{0i}$
is {\it not}. We will return to this point in Section~\ref{sec:4}, in
connection
with the possible excited states of the theory.
Using Eq.(\ref{CaS}), the following commutators are obtained:
\begin{equation}
[P_0,M_{0i}]=ix_0 \int d^3x \:\partial_i A_0(x) G(x)+i \int d^3x \:A_i(x) G(x)
+iP_i, \hspace{0.25in}[P_0,M_{kj}]=0,
\label{pomoi}
\end{equation}
\begin{equation}
[P_j,M_{0i}]=-i\delta_{ij}P_0, \hspace{1in} [P_i,M_{kj}]=i\delta_{i[k}P_{j]},
\label{pjmoi}
\end{equation}
\begin{equation}
[M_{0i},M_{0j}]=-iM_{ij},
\label{moimoj}
\end{equation}
\begin{equation}
[M_{kj},M_{lm}]=i\delta_{k[l}M_{m]j}+i\delta_{j[l}M_{k|m]}.
\end{equation}
Together with (\ref{PiPi}),
we thus recover the Poincar\'e sector of
(\ref{algebra}).

Turning to the twisted supersymmetries, we have the scalar charge Q,
identified in \cite{Witten}. It is preserved on an arbitrary manifold
and the energy momentum tensor can by
expressed as a $Q$ variation. In the context of the BRST construction of
(\ref{RLa}), it is precisely the BRST charge. Its expression is:
\begin{equation}
Q=\int d^3 x\: \biggl\{ \left( F_{0i}+ \tilde{F}_{0i}\right) \psi_i-\eta
D_0\phi-D_i\phi\:\chi_i-{\psi_0\over 2}\left[\lambda,\phi\right]\biggl\}.
\label{Q}
\end{equation}
Under translation and rotation, it transforms as:
\begin{equation}
[P_0,Q]=0=[P_j,Q],
\end{equation}
\begin{equation}
[M_{0i},Q]=i\int d^3x\:x_i\psi_0(x)G(x),\hspace{1in} [M_{kj},Q]=0,
\label{Pi5}
\end{equation}
\begin{equation}
{1\over 2}\{Q,Q\}=-\int d^3x\:\phi(x)G(x).
\end{equation}
We thus recover the nilpotency of $Q$ (up to gauge transformations),
but (\ref{Pi5}) shows that
$M_{\alpha\beta}$ does not correspond to the generator $J_{\alpha\beta}$
appearing in (\ref{algebra}).
This is confirmed by the study of $\bar{Q}_\alpha$, also identified in
\cite{Witten}.
Its time and space components are given by:
\begin{equation}
 \bar{Q} \equiv  \bar{Q}_0=\int d^3 x\: \biggl\{ \left( F_{0i}-
\tilde{F}_{0i}\right) \chi_i+\psi_0 D_0\lambda-\psi_iD_i\lambda
+{\eta\over 2}\left[\phi,\lambda\right]\biggl\},
\label{Qbar}
\end{equation}
\begin{equation}
\bar{Q}_i=\int d^3 x\: \biggl\{ \varepsilon_{ijk}\left( F_{0j}-
\tilde{F}_{0j}\right) \chi_k+\psi_0 D_i\lambda+\psi_i D_0\lambda
-\left( F_{0i}-\tilde{F}_{0i}\right) \eta+\varepsilon_{ijk}\psi_j D_k\lambda
+{1\over 2}\left[\phi,\lambda\right]\chi_i\biggl\}.
\label{Qi}
\end{equation}
The spacetime symmetry transformations of $\bar{Q}$ and $\bar{Q}_i$ are
made clear by:
\begin{equation}
[P_0,\bar{Q}]=-i\int d^3x\:\eta(x)G(x),\hspace{1in} [P_i,\bar{Q}]=0,
\end{equation}
\begin{equation}
[M_{0i},\bar{Q}]=i\bar{Q}_i-i\int d^3x\:x_i\eta(x)G(x),
\hspace{1in} [M_{kj},\bar{Q}]=0,
\label{moiq}
\end{equation}
\begin{equation}
{1\over 2}\{\bar{Q},\bar{Q}\}=-\int d^3x\:\lambda(x)G(x),
\end{equation}
as well as:
\begin{equation}
[P_0,\bar{Q}_i]=-i\int d^3x\:\chi_i(x)G(x),\hspace{1in} [P_j,\bar{Q}_i]=0,
\label{Qibar}
\end{equation}
\begin{equation}
[M_{0i},\bar{Q}_j]=i\delta_{ij}\bar{Q},\hspace{1in}
[M_{kj},\bar{Q}_i]=i\delta_{i[j}\bar{Q}_{k]},
\end{equation}
\begin{equation}
{1\over 2}\{\bar{Q}_i,\bar{Q}_j\}=-\delta_{ij} \int d^3x\: \lambda(x) G(x),
\hspace{1in}{1\over 2}\{\bar{Q},\bar{Q}_i\}=0
\end{equation}
When $\bar{Q}_\alpha$ is anticommuted with the BRST generator, it gives:
\begin{equation}
{1\over 2}\{ Q,\bar{Q}_\alpha \}=-P_\alpha-\int d^3x\: A_\alpha(x) \, G(x).
\label{Qalpha}
\end{equation}
Thus, given our choice of generators, (\ref{Q}) (\ref{Qbar}) and  (\ref{Qi}),
the relation (\ref{algg}) is obtained, provided $a_1 a_2=\frac{1}{2}$.
Observe how the adjoint assignments (\ref{adj}) produce:
\begin{equation}
 Q^ \dagger=- \bar{Q},
\label{Qdag}
\end{equation}
 and as announced they render the Hamiltonian (\ref{P0}) semi-positive
definite\ (as is the case in SYM).

To identify $S_{\mu\nu}$, we compute the adjoint of $\bar{Q}_i$, obtaining:
\begin{equation}
S_{0i}=\int d^3x\: \left[ \varepsilon_{ijk}\left( F_{0j}+\tilde{F}_{0j} \right)
\psi_k +\eta D_i \phi -\chi_i D_0\phi-\left( F_{0i}+\tilde{F}_{0i} \right)
\psi_0 + \varepsilon_{ijk}\chi_j D_k \phi -{1\over 2} \left[
\lambda, \phi \right] \psi_i \right] .
\label{S0i}
\end{equation}
Its spacetime symmetry transformations and nilpotency are
revealed by the following set
of commutators:
\begin{equation}
[P_0,S_{0i}]=-i\int d^3x\:\psi_i(x)G(x),\hspace{1in}[P_j,S_{0i}]=0,
\label{P0s0i}
\end{equation}
\begin{equation}
[M_{0i},S_{0j}]=-i\varepsilon_{ijk}S_{0k},\hspace{1in}[M_{kj},S_{0i}]=
i\delta_{i[j}S_{0|k]},
\label{soim}
\end{equation}
\begin{equation}
{1\over 2}\{S_{0i},S_{0j}\}=-\delta_{ij}\int d^3x\:\phi(x)G(x).
\end{equation}
Eqs.(\ref{P0s0i}) and (\ref{soim}) show that $S_{0i}$ generates
a symmetry if Gauss's law is imposed and that it is a self-dual object.
Relating to previous fermionic symmetries, we compute:
\begin{equation}
\{S_{0i},Q\}=0,
\end{equation}
\begin{equation}
{1\over 2}\{S_{0i},\bar{Q}\}=P_i+\int d^3x\:A_i(x)G(x),
\label{soiq}
\end{equation}
\begin{equation}
{1\over 2}\{S_{0i},\bar{Q}_j\}=-\delta_{ij}\left(P_0+
\int d^3x\:A_0(x)G(x)\right)+\varepsilon_{ijk}\left(P_k+\int d^3x\:A_k(x)G(x)
\right),
\label{soiqj}
\end{equation}
which reproduces (\ref{algh}), if $a_1 a_3 = \frac{-i}{4}$.

Noting now that the boost generators are not hermitian, we extract from them
the twisted internal generators by taking the anti-hermitian
part $R_{0i}\equiv M^\dagger_{0i}
-M_{0i}$. In terms of the fields, it is simply:
\begin{equation}
R_{0i}=\int d^3x\: \left( -i \psi_i \eta + i \psi_0\chi_i +i \varepsilon_{ilm}
\psi_l \chi_m \right).
\label{R0i}
\end{equation}
Commuting with the Poincar\'e generators produces:
\begin{equation}
[P_0,R_{0i}]=0, \hspace{1in}[P_j,R_{0i}]=0,
\end{equation}
\begin{equation}
[M_{0i},R_{0j}]=-i\varepsilon_{ijk}R_{0k},\hspace{1in}[M_{kj},R_{0i}]=
i\delta_{i[j}R_{0|k]},
\end{equation}
which shows that $R_{0i}$ is also a self-dual object.
When commuted with the fermionic symmetries and with itself, we get:
\begin{equation}
[R_{0i},Q]=iS_{0i},\hspace{1in}[R_{0i},\bar{Q}]=i\bar{Q}_i,
\end{equation}
\begin{equation}
[R_{0i},\bar{Q}_j]=i\varepsilon_{ijk}\bar{Q}_k
-i\delta_{ij}\bar{Q},\hspace{1in}[R_{0i},S_{0j}]=i\varepsilon_{ijk}S_{0k}
-i\delta_{ij}Q,
\end{equation}
\begin{equation}
[R_{0i},R_{0j}]=2i \varepsilon_{ijk} R_{0k}.
\label{roiroj}
\end{equation}
We thus find that (\ref{algebra}) is realized in TYM with the
following values of parameters:
$a_1 = 1,~a_2=\frac{1}{2},~a_3=\frac{-i}{4}$ and $a_4=\frac{1}{8}$.
As an infinitesimal transformation, $R_{\alpha\beta}$ only acts on
fermionic fields (as is obvious from (\ref{R0i}) and in parallel with SYM) :
\begin{eqnarray}
\delta_R\eta &=& {1 \over 2} \zeta_{\rho\sigma} \chi^{\rho\sigma},\nonumber \\
\delta_R \psi_\alpha &=& -2\zeta_{\alpha\lambda} \psi^{\lambda},\nonumber \\
\delta_R\chi_{\alpha\beta} &=& 2\zeta_{\alpha\beta}\eta-\zeta_{\lambda[\alpha}
\chi_{\beta]}{}^\lambda,
\end{eqnarray}
where $\zeta_{\alpha\beta}$ is an infinitesimal, commuting and self-dual
parameter.
Although relatively simple, this symmetry appears to have escaped notice.  It
would be interesting to investigate its use, for instance, in the
perturbative renormalization of TYM \cite{Brooks,Marcu} or determine the class
of
manifolds on
which it is preserved \cite{Ogievetsky}.

So far, we have thus identified for TYM all the generators in the twisted
$N=2$ superalgebra (\ref{algebra}), with the exception of $J_{\alpha\beta}$.
This generator should be hermitian, since it is so before twisting.
The more or less natural object to consider here is the hermitian part of
$M_{0i}$. So we conjecture:
\begin{mathletters}
\label{joikj}
\begin{eqnarray}
J_{0i} &=& M_{0i} + \frac{R_{0i}}{2}, \\
J_{kj} &=& M_{kj} + \frac{1}{2} \varepsilon_{kjl} R_{0l}, \label{Jkj}
\end{eqnarray}
\end{mathletters}
where in (\ref{Jkj}), we have used the self-duality of $R_{\mu\nu}$.
Using the relations previously obtained, we find that on physical states,
(\ref{alga} --- \ref{algb}),
(\ref{algd} --- \ref{algf}), and
(\ref{algn})
are verified, with the above mentioned values  of $a_i$'s.
Thus (\ref{joikj}) is indeed the correct identification.
In fact, this relation should be expected.
After twisting, the Lorentz algebra is isomorphic to
$SU_{L'}(2) \bigotimes SU_R(2)$ and thus some hybridization of the
internal symmetry with the old Lorentz generators $J_{\alpha\beta}$ is
expected.

\section{HERMITICITY AND EXCITED STATES}
\label{sec:4}

As shown in the last section, TYM theory in flat Euclidian
spacetime realizes the
$SO(4)$ ``Lorentz'' algebra in such a way that the boost generators $M_{0i}$
are
non-hermitian (neither are they antihermitian).
In order to classify the possible states of the theory, we wish to identify
the unitary representations of the symmetry algebra.  Let us concentrate here
on the compact sector $SO(4)$.  As is well known, the irreducible and unitary
representations are in that case finite dimensional
(dimension $(2\ell_1+1)(2\ell_2+1)$
with $\ell_1,\ell_2=0,1,2\ldots$),
the generators are represented by hermitian matrices and the group
elements related
to the identity can be written as $e^{i \, {\cal \alpha}_{\mu\nu} \,
M^{\mu\nu}}$
with parameters $\alpha_{\mu\nu}$.
Now if $M_{0i}$ is not hermitian, it is clear that as far as $SO(4)$ is
concerned, the only admissible unitary representation will be the trivial one,
in which
$M_{0i}=0$.  The $SO(3)$ subgroup of spatial rotations generated by $M_{kj}$
does not suffer this problem, and the Hilbert space of the theory could carry
the usual labels $\ell m$ of the $SO(3)$ representation since this subgroup
commutes with $P_0$.  But because $M_{0i}$ are not hermitian, only $\ell=m=0$
will be present in that case.  This can be seen in the algebra: acting with
both sides of (\ref{moimoj}) on the representation space
will give the same result provided
$M_{ij}$ is also vanishing.

If TYM is considered in Minkowski spacetime,
with $g_{\mu\nu}=\eta_{\mu\nu}$, $M_{0i}$ will also  be non-hermitian,
with equally dramatic consequences.
Suppose we are interested in the unitary
representations of the twisted algebra (\ref{algebra}), assumed to be
rotated to Minkowski metric, as specified in section \ref{sec:3}.
To investigate them,
we make use, as in the case of the superPoincar\'e algebra \cite{West}, of
Wigner's method of induced representations \cite{Wigner}.  This method is also
appropriate here since our symmetry group possesses the same abelian invariant
subgroup, namely the translations.
In this method, one first makes a choice of  ``standard vector'',
eigenstates of $P_\mu$ and a representative member of
the possible classes of eigenvalues of the
Casimir $P_\mu^{\,2}$.
One then identifies the little group, formed by the generators that leave the
standard vector intact, and excluding the abelian subgroup.
Once the irreducible unitary representations of the little group have been
identified (restricting to finite dimensional ones), they are then used to
induce
an  irreducible unitary representation of the whole group.
This is done by acting on the standard vector with the generators that
change its eigenvalue of $P_\mu$.
These infinite dimensional representations then form the plane-wave basis, to
which particles are associated.

Consider the massless case. The little supergroup is formed by
$C_1 \equiv M_{1 0} + M_{1 3}, C_2 \equiv M_{2 0} + M_{2 3}$,
$M_{1 2},  Q, \bar{Q}_\alpha, S_{\mu\nu}, R_{\mu\nu}$.
Acting with any of these will leave the vector
$\left| \, p_0^{\,\mu} = (m,0,0,m) \, \right\rangle$ unrotated.  Now since
\begin{eqnarray}
\relax [C_1,~M_{1 2}] &=& -C_2, \nonumber \\
\relax [C_2,~M_{1 2}] &=&  C_1, \nonumber \\
\relax [C_1,~C_2]     &=&  0,
\end{eqnarray}
is the Lie algebra $E_2$, and since we seek a finite dimensional
representation, we are led to $C_1=C_2=0$ when acting on the standard vector,
just as in the superPoincar\'e algebra \cite{West}.
Thus, at this level, the non-hermiticity of $M_{1 0}$ and $M_{2 0}$ appears
irrelevant.
However, in order to induce a representation of the entire group, we need a
unitary realization of the finite transformation generated by
$M_{3 0}$,
$M_{1 0}$ -- $M_{1 3}$ and
$M_{2 0}$ -- $M_{2 3}$.
But with $M_{3 0}$ non-hermitian, this can only be implemented
through a trivial realization: $M_{0 3}=0 $.
This in turn implies that if we consider the first part of Eq.~(\ref{pomoi})
and
choose $i=3$ when acting on $\left| \, p_0 \, \right\rangle$,
the LHS will vanish, and lead to $P_3 | p_0 \rangle = 0$. (We refer to
euclidean commutators for convenience; at this point the results clearly
do not depend on the signs  appearing in them.)
One thus conclude that massless excitations will not occur in TYM.

A similar situation occurs if one attempts to construct massive
representations.  Taking as the standard vector
$| p_1^{\,\mu}=(m,0,0,0) \rangle$, the little group is made of
$(M_{kj}, J_{kj}, Q, \bar{Q}_\alpha, S_{\mu\nu}, R_{\mu\nu})$.
Inducing a representation of the whole group will require a unitary operator
for finite boosts, again this is only possible if the action of $M_{0i}$ is
trivial:
$M_{0i} | p_1 \rangle = 0$.
But using now the first part of (\ref{pjmoi}), we find
$P_0 | p_1 \rangle = 0$, again contradicting the assumption  on $| p_1
\rangle$.
In this way, we recover, in a group theoretical context,
the absence of dynamics in TYM.

We now focus on the last possibility : null representations with standard
vector $| p_3^{\,\mu} = 0 \rangle$. (We will not consider spacelike
representations). This vector is left unchanged by any
Lorentz transformation and the little group is made of all the generators:
$M_{\alpha\beta}$,
$J_{\alpha\beta}$,
$Q$,
$\bar{Q}_\alpha$,
$S_{\alpha\beta}$ and $R_{\alpha\beta}$.
Here, representations of the full group and the little group coincide.
As before, because we seek unitary representations, we will require that
$M_{0i} | p_3 \rangle = 0$.
When used in (\ref{moimoj}) we obtain $M_{kj} | p_3 \rangle = 0$,
showing the rotational invariance of $| p_3 \rangle$, which has thus
the characteristics of a vacuum state.
Turning now to the action of $Q$, consider the time component of
(\ref{Qalpha}), it reads:
\begin{equation}
\{ Q,Q^{\dagger} \} \, | p_3 \, \rangle = 0
\label{aaa}
\end{equation}
since $|p_3\rangle$ is by construction a physical state.  Projecting on
$\langle p_3 |$,
we find
\begin{equation}
\langle Q \, p_3 | Q \, p_3 \rangle +
\langle Q^{\dagger} \, p_3 | Q^{\dagger} p_3 \rangle = 0
\label{bbb}
\end{equation}
and conclude that $Q|p_3\rangle=Q^{\dagger}|p_3\rangle=0$.

Similarly, we can easily determine that the other generators have
eigenvalue 0.  By (\ref{Qdag}), $\bar{Q} | p_3 \rangle = 0$.
Making use of (\ref{moiq}), we then find
$\bar{Q}_i | p_3 \rangle = 0$.
Applying the same reasoning with (\ref{soiqj}) and (\ref{roiroj}), we
find $S_{0i}|p_3\rangle=0$ and $R_{0i}|p_3\rangle=0$.
Thus, all generators act trivially in TYM.

Now as mentioned before, the Lagrangian in (\ref{RLa}) can be obtained by gauge
fixing of a topological symmetry .  The BRST charge introduced in that
construction is the scalar $Q$ given in (\ref{Q}).  In that context, the
physical states are assumed to be annihilated by $Q$, and such that they are
not of the form $Q|\alpha\rangle$.  Having shown the former, we now argue for
the latter, following Ref.~\cite{Witten}.
Consider a state $|\psi\rangle=Q|\alpha\rangle$,
with $P_0|\psi\rangle=0$.
Because $[P_0, Q]=0$, $|\psi\rangle$ and $|\alpha\rangle$
can be chosen to have the same eigenvalue under  $P_0$.
But with $P_0 | \alpha \rangle = 0$, applying the steps given in (\ref{aaa})
and (\ref{bbb})
will lead to $|\psi\rangle=0$.
We thus obtain, in the context of twisted $N=2$ SYM, the
BRST cohomology condition of Refs.~~\cite{Witten,Brooks,Singer}
on physical states.

\section{CONCLUSION}
\label{sec:5}
We have used the Hamiltonian formalism
to study the symmetries  of (\ref{RLa}). This formalism  offers the
inconvenience of a non manifest covariance, but made explicit the generators,
as well as the ``propagation'' of the Gauss law constraint through the algebra.
In this context, it would be interesting to see how the algebra we have
obtained is modified by the gauge fixing of the Yang-Mills symmetry
\cite{OSB}.
We were also able to make precise the relation between the Lorentz generators
of TYM ($M_{\alpha\beta}$) and the twisted version of Lorentz and $SU_I(2)$
generators of SYM $(J_{\alpha\beta}$ and $R_{\alpha\beta}$ respectively)
as  displayed in
(\ref{joikj}). It is usually not illuminating to add symmetries to obtain
new ones, but the interest here lies in their physical significance. One could
avoid introducing the non-hermitian $M_{0i}$. But in order to
understand the Lorentz structure of the
various objects (fields, charges, etc ) of the theory, they are needed.
It is thus more sensible
to discard $J_{\alpha\beta}$, keeping $M_{\alpha\beta}$ and $R_{\alpha\beta}$.
In this way, $R_{\alpha\beta}$ appears as a symmetry of
(\ref{RLa}) unappreciated in previous work. In fact,
its existence may seem odd at first
sight, in view of the Coleman-Mandula theorem\cite{ColeMan}.
But as we have shown in
Section~\ref{sec:4}, no massive unitary representations are realised in TYM,
and
in this way, the conclusions of the theorem are inapplicable.
Nevertheless, more could be learned about  $R_{\alpha\beta}$.
Extending to more general manifolds, is it preserved\cite{Ogievetsky}?
Can it be used,
along the lines of \cite{Oliveira}, to draw conclusions on the quantum theory
at all orders in perturbation theory by restricting the possible counterterms
(provided anomalies are absent)?
It would also be interesting to investigate the extent of that symmetry
in other topological theories. For instance, the symmetry algebra of
the Chern-Simons theory in the Laudau
gauge has been found to coincide with  a twisted $N=4$ superalgebra
\cite{DGS,BDL}.
It is expected that a  twisted internal symmetry will also exist in that case.

\acknowledgments
I thank Roger Brooks for suggesting this investigation, and providing direction
along the way. I also benefitted from comments from Daniel Freedman, Roman
Jackiw, Kenneth Johnson, Choonkyu Lee and Claudio Lucchesi.

\end{document}